\documentclass[10pt,conf]{IEEEtran}
\ifCLASSINFOpdf \else \fi
\usepackage{geometry}
\usepackage[normalem]{ulem}
\usepackage{multirow}
\usepackage{bbold}
\usepackage{ifpdf}
\usepackage{ulem} 
\usepackage{subcaption}%
\usepackage[cmex10]{amsmath}
\usepackage{amssymb}
\usepackage{verbatim}
\usepackage{algorithm}
\usepackage{algorithmic}
\usepackage{array}
\usepackage{latexsym}
\usepackage{color}
\usepackage{textgreek}
\usepackage{subscript}
\usepackage{rotating}
\usepackage{booktabs,multirow}
\usepackage{mdframed}	
\usepackage{tabularx}
\usepackage{amsmath}
\usepackage{makecell}
\usepackage{tabto}
\usepackage{mathrsfs}
\usepackage{url}
\usepackage{cite}
\usepackage{listings}
\usepackage{array}
\usepackage[table]{xcolor}
\usepackage{ragged2e}
\usepackage{booktabs}

\usepackage{graphicx}
\usepackage{comment}
\usepackage[cmex10]{amsmath}
\usepackage{amssymb}
\usepackage{bbold}
\usepackage{float}
\usepackage[cmex10]{amsmath}
\usepackage{amssymb}
\usepackage{verbatim}
\usepackage{array}
\usepackage{latexsym}
\usepackage{color}
\usepackage{tabularx} 
\usepackage{textgreek}
\usepackage{subscript}
\usepackage{rotating}
\usepackage{booktabs,multirow}
\usepackage{mdframed}	
\usepackage{tabularx}
\usepackage{amsmath}
\usepackage{makecell}
\usepackage{tabto}
\usepackage{mathrsfs}
\usepackage{url}
\usepackage{cite}
\usepackage{listings}
\usepackage{array}
\usepackage[table]{xcolor}

\newcommand{\noteblue}[1]{\textcolor{blue}{{\bf #1}}}
\definecolor{pinkpurple}{rgb}{0.6, 0.1, 0.9} 
\usepackage{algorithmic}
\usepackage{array}
\usepackage{soul}
\sethlcolor{green}
\usepackage{url}
\usepackage{pifont}
\usepackage{xcolor}
\usepackage{multirow}
\usepackage{graphicx} 

\begin{document}

\title{LiLM-RDB-SFC: Lightweight Language Model with Relational Database-Guided DRL for Optimized SFC Provisioning}

\author{
\IEEEauthorblockN{Parisa~Fard~Moshiri$^1$, Xinyu Zhu$^1$, Poonam Lohan$^1$, Burak Kantarci$^1$, Emil Janulewicz$^2$,}\\
\IEEEauthorblockA{\textit{$^1$University of Ottawa, Ottawa, ON, Canada}\\
\textit{$^2$Ciena, 383 Terry Fox Dr,
Kanata, ON K2K 2P5, Canada}\\
$^1$\{parisa.fard.moshiri, xzhu095, ppoonam, burak.kantarci\}@uottawa.ca,~$^2$ejanulew@ciena.com}
}

\maketitle
\thispagestyle{empty}
\pagestyle{empty}
\begin{abstract}
Effective management of Service Function Chains (SFCs) and optimal Virtual Network Function (VNF) placement are critical challenges in modern Software-Defined Networking (SDN) and Network Function Virtualization (NFV) environments. Although Deep Reinforcement Learning (DRL) is widely adopted for dynamic network decision-making, its inherent dependency on structured data and fixed action rules often limits adaptability and responsiveness, particularly under unpredictable network conditions. This paper introduces LiLM-RDB-SFC, a novel approach combining Lightweight Language Model (LiLM) with Relational Database (RDB) to answer network state queries to guide DRL model for efficient SFC provisioning. Our proposed approach leverages two LiLMs,  Bidirectional and Auto-Regressive Transformers (BART) and the Fine-tuned Language Net T5 (FLAN-T5), to interpret network data and support diverse query types related to SFC demands, data center resources, and VNF availability.  Results demonstrate that FLAN-T5 outperforms BART with a lower test loss (0.00161 compared to 0.00734), higher accuracy (94.79\% compared to 80.2\%), and less processing time (2h 2min compared to 2h 38min). Moreover, when compared to the large language model SQLCoder, FLAN-T5 matches the accuracy of SQLCoder while cutting processing time by  96 \% (SQLCoder: 54 h 43 min; FLAN-T5: 2 h 2 min).

\end{abstract}

\begin{IEEEkeywords} SFC provisioning, VNF placement, DRL, Language Model, FLAN-T5, BART, Network State Monitoring.
 
\end{IEEEkeywords}

\IEEEpeerreviewmaketitle

\section{Introduction}
The advent of Software-Defined Networking (SDN) and Network Function Virtualization (NFV) has altered network management, allowing for greater flexibility and efficiency. Service Function Chain (SFC) provisioning involves the sequential execution of different virtual network functions (VNF) to support complex applications such as Cloud Gaming (CG), Augmented Reality (AR), Video Streaming (VS),  Massive IoT (MIoT), Voice over Internet Protocol (VoIP), and Industry 4.0 (Ind 4.0)\cite{han2024}. Satisfying these applications through SFC provisioning presents substantial challenges, such as resource allocation for VNFs placements, sequential VNF execution, and meeting end-to-end (E2E) latency constraints. 

Deep Reinforcement Learning (DRL) algorithms are frequently employed for optimal VNF placement and SFC provisioning due to their ability to handle complex network environments, make effective resource allocation decisions, and adapt to varying service demands\cite{mohamed2024}. However, DRL techniques often rely on structured numerical inputs, predetermined state-action representations, and learned policies, restricting their adaptability while encountering unexpected network states or scenarios that significantly differ from their training phase.
For instance, when typical data flows are interrupted by unexpected network outages or link failures, a DRL model trained on stable and structured network conditions may struggle to swiftly reroute traffic or reposition VNFs effectively in such a situation, maintaining the placement decisions or routing paths learned from regular operation. This could result in additional outages, higher latency, and poor network performance \cite{parisa2025}.

In contrast, integrating Language Models (LMs) can significantly enhance decision-making by exploiting their abilities to comprehend unstructured input, reason about context, and quickly adapt to unexpected network scenarios \cite{hoang23}. LMs can scan real-time textual descriptions or logs that specify the nature of failures, allowing for faster contextual evaluation and dynamic decision-making beyond the established numerical metrics and structured state-action restrictions inherent in DRL \cite{boateng2024}. Specifically, LMs address this issue by immediately evaluating the severity of partial outages or link failures from unstructured event data, allowing them to quickly inform the issue and suggest alternative routing paths or ideal VNF relocation strategies \cite{parisa2025}. 
Thus, combining DRL and LMs enables more informed, adaptive, and responsive SFC provisioning and VNF placement decisions, minimizing the adverse effects caused by unexpected network events.

In our previous work \cite{arda24}, VNF placement using DRL is addressed, focusing on the optimal allocation of storage and computational resources required by VNFs. The goal in \cite{arda24} is to maximize the efficient handling of SFC requests based on resource constraints and VNF specifications. Initially, our approach in \cite{parisa2025} involves using a textual dataset to capture network states information; however, in this work, we transition to a relational SQL database due to its superior capability for handling scalability and facilitating rapid reconfigurability across diverse DCs, varied VNFs and SFC scenarios. After storing network state information determined by DRL actions in a relational database, the dataset is fed into a Lightweight LM (LiLM) for comprehensive network state monitoring.  This integration offers deep and actionable insights into network dynamics, allowing greater adaptability and decision-making precision in DRL-based VNF placement. LiLMs require substantially fewer computational resources and less advanced hardware compared to Large LMs (LLM), achieving comparable performance while significantly improving inference speed and operational efficiency \cite{hassid2024}. Their suitability for rapid, cost-effective decision-making makes them highly valuable for dynamic network optimization tasks \cite{boateng2024}. 

In this paper, we employ both BART and FLAN-T5, two lightweight yet powerful language models, to translate natural-language query to SQL queries. BART is known for its strong performance in sequence-to-sequence tasks, particularly text generation and summarization, while FLAN-T5 stands out for its instruction tuning and generalization capabilities across diverse natural language (NL) processing tasks. Both models are assessed across diverse question types centered on Data Center (DC), VNF, and SFC configurations. To benchmark the performance of LiLMs, we also employ a state-of-the-art LLM, SQLCoder, which is specifically designed for efficient SQL query generation. The results highlight the LiLMs capability to detect resource usage patterns, pinpoint performance bottlenecks, and extract valuable insights for future network enhancements  with lower computational cost compared to the LLM. Consequently, the system evolves to be more adaptive and intelligent, enabling proactive handling of SFC provisioning in dynamic environments.

The main contributions of this paper are as follows: 
\begin{enumerate}
\item We design and implement a structured relational SQL database schema that captures essential network state information, including storage and computational resources of DCs, available idle VNFs, and current SFC demands. Additionally, a custom dataset comprising natural-language and SQL query pairs is curated to train the proposed language models for SQL query generation.

\item Two LiLMs, BART and FLAN-T5, and a LLM, SQLCoder, are trained on the custom dataset to translate diverse natural-language queries into corresponding SQL queries. These SQL queries are executed over the relational SQL database to retrieve accurate, real-time network state information.

\item The proposed system demonstrates high scalability and adaptability to dynamic network conditions. The relational database schema can be extended by adding new rows, while the trained LiLMs generalize well to unseen queries and scenarios without additional training.
\end{enumerate}

Based on our findings, FLAN-T5 outperforms BART with lower test loss, higher accuracy, higher number of correct predictions, and less processing time. Compared to SQLCoder, FLAN-T5 has almost same accuracy but with lower processing time. The rest of the paper is organized as follows: Section II provides a literature review, followed by the methodology in Section III. Section IV discusses performance evaluation. Section V concludes the paper.

\section{Related Work}
In NFV systems, deep learning methodologies have been investigated to enhance predictive precision for VNF resource management and service orchestration. Kim et al. \cite{kim2019deep} present a sequence modeling framework for VNF resource prediction utilizing LSTM (Long short-term memory) variations and attention processes. Their strategy enhances prediction accuracy and convergence speed by utilizing the structural interdependence inherent in SFCs. Nonetheless, these methodologies generally depend on supervised training using structured time-series data and exhibit limited adaptation to dynamic and unstructured network contexts. Bunyakitanon et al. \cite{bunyakitanon2020arel3p} introduce AREL3P, a RL framework employing Q-learning to independently position VNFs based on predictions of E2E performance. In contrast to supervised learning methods, AREL3P interfaces with NFV orchestration platforms and facilitates online learning in dynamic contexts. Yet, as a tabular RL method, it encounters scalability constraints and lacks semantic interpretability. Although both supervised and non-DRL approaches possess distinct advantages, they are hindered by inflexible input representations and exhibit restricted adaptability in managing unstructured or dynamic network environments.

DRL has been extensively utilized to tackle dynamic SFC provisioning and VNF placement. Fu et al. \cite{fu2020dynamic} present a method based on DRL for the embedding of VNFs in NFV-enabled IoT systems, disaggregating VNFs into granular functional components to enhance flexibility. Their Deep Q-Learning approach, augmented with experience replay and target networks, proficiently tackles traffic fluctuation and infrastructure heterogeneity. Jaumard et al. \cite{jaumard2024} integrate DRL with Graph Neural Networks (GNN) to encapsulate topological and functional restrictions in SFC routing decisions. While effective, these models rely on predetermined state-action representations and provide limited interpretability in the presence of unstructured or unforeseen network alterations. Our methodology enhances DRL by incorporating a streamlined LM module that facilitates semantic querying of DRL-generated judgments within a structured SQL framework.

Recent studies have investigated the application of LMs in network design, forecasting, and decision-making support. Su et al. \cite{su2024llm} utilize pre-trained LLaMA2 models for zero-shot prediction of VNF resource utilization. By tokenizing numerical resource measurements and utilizing the sequence modeling capabilities of LMs, the model attains competitive accuracy without the need for task-specific fine-tuning. Nonetheless, it is constrained in organized reasoning and provides minimal interpretability for decision-making in operational settings. Nguyen et al. \cite{nguyen2024nfvintent} provide NFV-Intent, a framework that uses in-context learning to convert user intents into JSON-formatted NFV configurations. Their technology attains excellent translation accuracy and seamlessly interacts with the complete NFV lifecycle without necessitating fine-tuning. Li et al. \cite{li2025nextgen} introduce LM-NOS, utilizing LMs to enhance heuristic policy code inside a multi-objective optimization framework for SFC deployment. While these solutions utilize LMs for deployment logic or configuration abstraction, they concentrate on pre-deployment phases and lack support for real-time semantic querying of dynamic network states. Conversely, our research utilizes LiLM (FLAN-T5, BART) and a LLM (SQLCoder) to convert natural-language inquiries into SQL queries pertaining to structured network monitoring data, facilitating transparent and contextually aware interpretation of judgments generated by DRL.

\section{Methodology}
In our previous work \cite{arda24}, DRL is utilized to maximize the number of accepted SFC requests while considering infrastructure constraints. Given the diverse resource demands, latency requirements, and unique VNF sequences of different SFC requests, DRL effectively learned optimal placement policies tailored to specific network conditions. However, DRL faces challenges in quickly adapting to unforeseen changes or incorrect decisions, often necessitating extensive retraining.
To address these limitations, we now incorporate LMs with an SQL-based dataset for managing SFCs, DCs, and VNFs. The types of SFCs and their VNF sequences used in this work are provided in \tablename\hspace{0.1pt}~\ref{tab:sfc_characteristic}. The primary VNFs of these SFCs can
be listed as Network Address Translation (NAT), Intrusion Detection and Prevention System (IDPS), Video Optimization Controller (VOC), Firewall (FW),  Traffic Monitor (TM), and WAN Optimizer (WO).
 The SQL dataset offers structured data organization, efficient querying capabilities, and robust relational integrity, enabling quick and precise access to relevant network information. We store network state information in SQL database following DRL actions at each time-step. Then, LM is leveraged to convert NL query to SQL query to monitor the network state accessing the SQL database. NL queries are fed into the LM to dynamically investigate critical network state metrics such as minimum and maximum E2E latency for specific SFC, the number of idle VNFs, and available storage and computational power at a particular DC. By deeply understanding network state information, LMs can produce valuable inputs/recommendations for DC selection function, the output of which provides inputs to the DRL model for optimal VNF placements aligned with service requests and real-time network status. These recommendations can be provided either periodically or on demand. 

\begin{figure*}[!t]
        \centering
        \includegraphics[width = 0.85\textwidth, trim=0cm 0cm 0cm 0cm,clip]{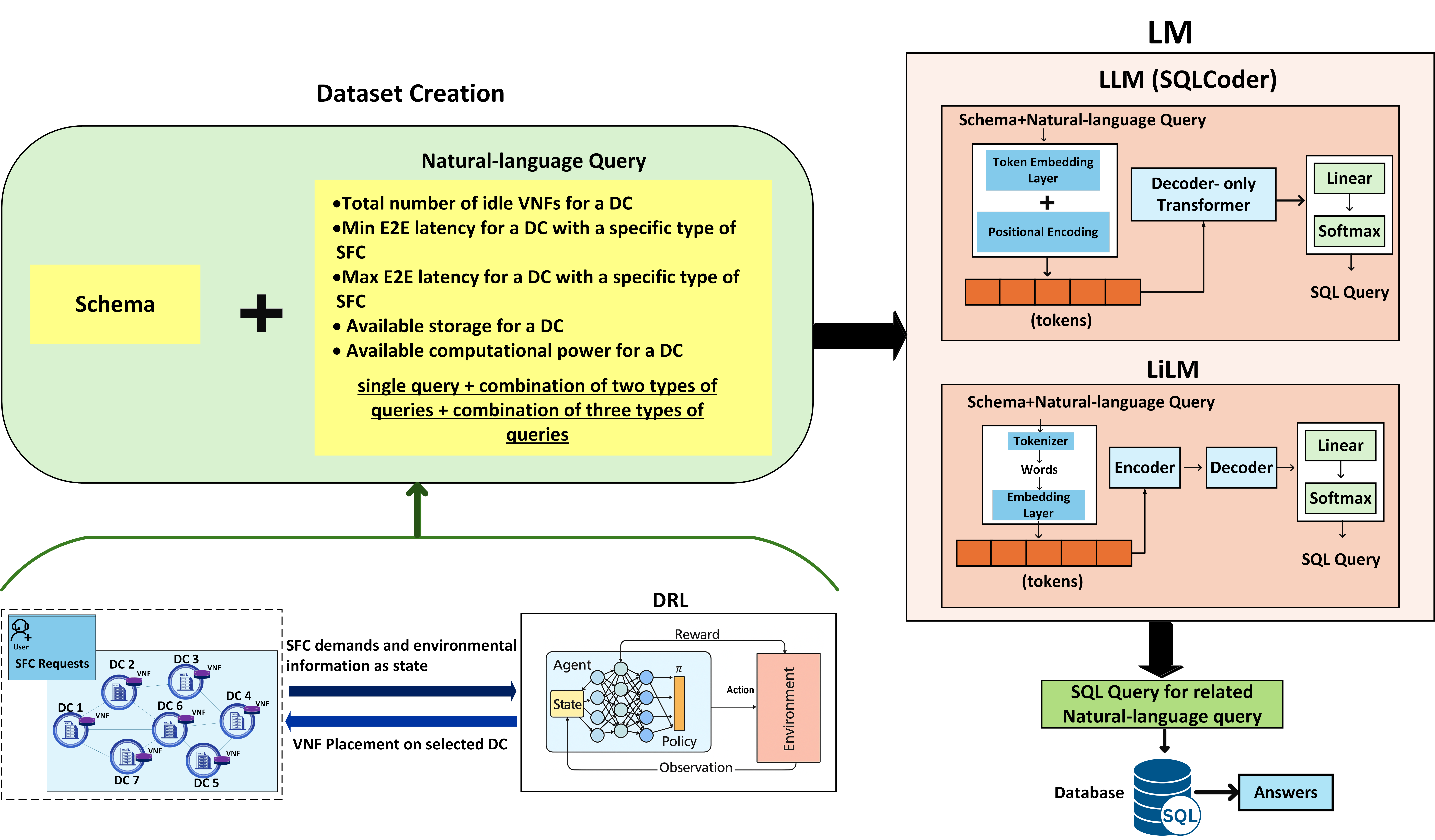}
        \caption{Framework Design for LiLM-RDB-SFC}
        \label{fig:system} 
\end{figure*}
\begin{figure*}[!t]
        \centering
        \includegraphics[width = 0.8\textwidth, trim=0cm 0cm 0cm 0cm,clip]{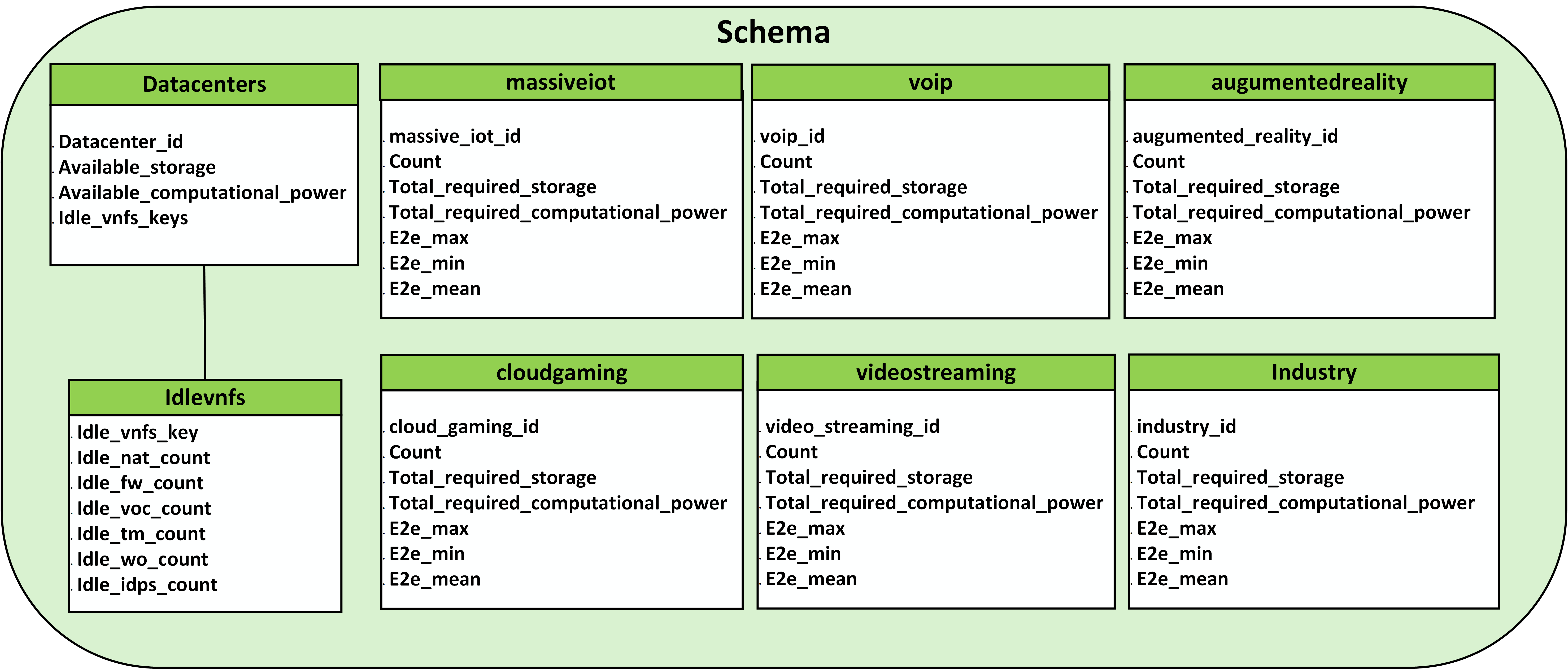}
        \caption{Schema for Relational Database}
        \label{fig:schema} 
\end{figure*}

As shown in \figurename\hspace{0.1pt} \ref{fig:system}, comprehensive network state information is collected following the actions taken by the DRL model. This data is then structured into a schema compatible with a relational SQL database, which is provided in details in \figurename\hspace{0.1pt} \ref{fig:schema}. The schema is fully dynamic, as individual rows can be updated on-the-fly, enabling real-time modification of any table entry without interrupting normal database operations. NL queries are formulated to retrieve key network metrics, focusing on: (i) the total number of idle VNFs, (ii) the minimum E2E latency for a specific SFC type at a given DC, (iii) the maximum E2E latency for a specific SFC type at a DC, (iv) available storage at a DC, and (v) available computational capacity at a DC. Furthermore, queries may involve combinations of two metrics (e.g., available storage and minimum E2E latency) and three metrics (e.g., minimum and maximum E2E latency along with the total number of idle VNFs), allowing for more detailed network state analysis.
These NL queries, along with the generated schema, are used for LM training. The LM is trained to translate NL queries into accurate SQL queries. When executed, these SQL queries retrieve relevant information from relational database, resulting in useful insights for network monitoring and decision-making processes in response to future network requests. Using this systematic technique, network management's responsiveness, accuracy, and adaptability can be substantially improved, allowing for more efficient and proactive resource allocation strategies tailored to dynamic network conditions.

\begin{table} 
\centering
\caption{SFC characteristics \cite{table}}\fontsize{6.5}{7.7}\selectfont
\begin{tabular}{|p{1.4cm}|p{1.6cm}|p{1cm}|p{1cm}| p{1 cm}|} 
 \hline
 \textbf{SFC Request}&\textbf{VNF Sequence} &\textbf{Bandwith (Mbps)} &\textbf{E2E delay (msec)} &\textbf{Request Bundle} \\  
 \hline
  CG & NAT-FW-VOC\break-WO-IDPS  & 4 & 80  & [40-55] \\
  \hline
  AR & NAT-FW-TM\break-VOC-IDPS & 100 & 10  & [1-4] \\
  \hline
  VoIP & NAT-FW-TM\break-FW-NAT & 0.064 & 100 & [100-200] \\
  \hline
  VS & NAT-FW-TM\break-VOC-IDPS & 4  & 100  & [50-100] \\
  \hline
  MIoT & NAT-FW-IDPS  & [1-50] & 5 & [10-15] \\
  \hline
  Ind 4.0 & NAT-FW & 70 & 8 & [1-4] \\
 \hline
\end{tabular}
\label{tab:sfc_characteristic}
\end{table}

\subsection{Integration of Language Models}

T5, which stands for ``Text-To-Text Transfer Transformer," is a flexible language model created by Google \cite{t5paper}. It efficiently treats every natural-language processing task as a text-to-text challenge \cite{t5paper}. During the pre-training phase, parts of the text are hidden, and the T5 model learns to fill in the gaps, which helps it develop a solid grasp of grammar, structure, and meaning. Fine-tuned Language Net T5 (FLAN-T5) is an improved version of T5 released by Google Research. It builds upon the original T5 architecture and was trained to follow natural-language instructions across a wide range of tasks, making it more effective in generalization and zero-shot scenarios.
In this research, FLAN-T5 is utilized for text-to-SQL generation. When given a natural-language query along with a description of the database schema, LiLM learns to create the SQL query that would provide the answer utilizing SQL database. Its ability to comprehend both the question's structure and the schema's relational format allows it to generate accurate SQL statements. 

BART (Bidirectional and Auto-Regressive Transformers) is also utilized as a complementary LiLM. BART, Developed by Facebook AI, is a powerful sequence-to-sequence architecture that integrates the strengths of BERT (bidirectional encoder) and GPT (autoregressive decoder) \cite{bartpaper}. Similar to T5, BART is pre-trained with a denoising autoencoder objective and can be fine-tuned for tasks like text-to-SQL generation. BART's adaptability and robustness make it a viable option for translating natural-language queries into executable SQL statements in network management. While both methods are effective for text-to-SQL creation, their relative efficacy may differ based on the dataset's features and the specific query patterns used. These patterns can include, but are not limited to, queries requiring aggregations or joins, value comparisons, or filtering based on certain attributes. Other significant linguistic or structural changes in the query could also affect model performance.

\textbf{SQLCoder}  is an open-source LLM specifically designed for translating natural language queries into SQL queries with high fidelity. Developed by Defog.ai \cite{defog-sqlcoder}, SQLCoder builds upon modern transformer architectures and is fine-tuned on diverse datasets consisting of complex text-to-SQL datasets \cite{defog2023sqlcoder}. Its core design adopts a decoder-only architecture, similar to models like LLaMA and GPT, leveraging multi-head self-attention and dense feedforward layers to capture intricate relationships between natural language input and SQL output \cite{sqlcoder}. To further enhance its adaptability and reduce fine-tuning resource requirements, SQLCoder can be updated using parameter-efficient techniques such as Low-Rank Adaptation (LoRA), which injects small, trainable rank-decomposition matrices into each attention and feed-forward block. This enables efficient domain adaptation without retraining the full model \cite{parisa2025}.
To enable the efficient deployment of LLMs such as SQLCoder on resource-constrained hardware, quantization techniques are commonly applied. 4-bit and 8-bit quantization reduce the precision of the model weights from 16 or 32-bit floating point values to lower-bit integer representations. This substantially decreases both the memory footprint and computational requirements, enabling faster inference and lower power consumption.
In practice, 8-bit quantization preserves much of the model’s original accuracy while offering notable efficiency gains, making it a standard choice for practical deployments. 4-bit quantization provides even greater compression and acceleration, allowing SQLCoder models to fit into devices with limited RAM and further reducing inference latency. Recent research and open-source libraries, such as bitsandbytes \cite{bitsandbytes2023}, have demonstrated that SQLCoder and similar LLMs can operate with minimal performance degradation when quantized to 4 or 8 bits, making advanced natural language-to-SQL capabilities accessible on a wide range of hardware platforms.

\newcolumntype{M}[1]{>{\centering\arraybackslash}m{#1}}

\begin{table}[!t]
\centering
\renewcommand{\arraystretch}{1.3} 

\caption{Notation Table}
\label{tab:notations}

\begin{tabularx}{0.45\textwidth}{|M{1.2cm}|X|M{0.95cm}|}
\hline
\textbf{Parameter} & \textbf{Description} & \textbf{Eq.} \\
\hline
\( i\) & each individual example in a batch & (\ref{eq:penalty-s}) - (\ref{eq:avg-penalty-iv})\\
\hline
\( N \) & the total number of examples in batch & (\ref{eq:avg-penalty-s}), (\ref{eq:avg-penalty-iv})\\
\hline
\( P_s^{(i)}\) &  binary penalty function defined for the \(i\)-th example in the batch for SFC identifiers & (\ref{eq:penalty-s}), (\ref{eq:avg-penalty-s})\\
\hline
\( P_{v}^{(i)} \) &  binary penalty function defined for the \(i\)-th example in the batch for idle VNFs identifiers &(\ref{eq:penalty-iv}), (\ref{eq:avg-penalty-iv})\\
\hline
\(s_i\) & expected SFC identifier for the \(i\)-th example &(\ref{eq:penalty-s})\\
\hline
\(\hat{s}_i\) & predicted SFC identifier for the \(i\)-th example &(\ref{eq:penalty-s})\\

\hline
\( {v}_i \) & expected Idle VNFs identifier for the \(i\)-th example&(\ref{eq:penalty-iv})\\
\hline
\(\hat{v}_i\) & predicted Idle VNFs identifier for the \(i\)-th example &(\ref{eq:penalty-iv})\\
\hline
\( P_S \) & average penalty regarding SFC identifiers&(\ref{eq:avg-penalty-s}), (\ref{eq:overall-loss})\\
\hline
\(P_{V}\) & average penalty regarding idle VNFs identifiers &(\ref{eq:avg-penalty-iv}), (\ref{eq:overall-loss})\\
\hline
\( \lambda_{ce} \) & penalty weight for cross-entropy loss&(\ref{eq:lambda-sum}), (\ref{eq:overall-loss})\\
\hline
\(\lambda_{s}\) & penalty weight for SFCs mismatch &(\ref{eq:lambda-sum}), (\ref{eq:overall-loss})\\
\hline
\( \lambda_{v}\) & penalty weight for idle VNFs mismatch&(\ref{eq:lambda-sum}), (\ref{eq:overall-loss})\\
\hline
\(L_{ce}\) & cross-entropy loss &(\ref{eq:overall-loss})\\
\hline
\(L\) & total loss &(\ref{eq:overall-loss})\\
\hline
\end{tabularx}
\end{table}
\subsection{Loss Function}
In order to improve the ability of LiLMs to capture the SFC and idle VNF identifiers correctly, additional penalty terms are introduced to the loss function. The standard cross-entropy loss may not explicitly enforce correct identification of these important components. By incorporating binary penalty functions for mismatches in predicting SFC and idle VNFs' identifiers, we can guide the model to focus on these details.  

For a batch of \(N\) examples, \(s_i\) is the expected SFC identifier and \(\hat{s}_i\) is the predicted SFC identifier for the \(i\)-th example. In addition, \(v_i\) is the expected idle VNF identifier and \(\hat{v}_i\) is the predicted idle VNFs identifier for the \(i\)-th example. The binary penalty functions $P_s^{(i)}$ and $P_v^{(i)}$ for SFC and VNF identifier, respectively, are defined as follows:
\vspace{+5mm}
\begin{align}
P_s^{(i)} =
\begin{cases}
1 & \text{if } s_i \neq \hat{s}_i \\
0 & \text{if } s_i = \hat{s}_i
\end{cases}
\label{eq:penalty-s}
\end{align}
\begin{align}
P_{v}^{(i)} =
\begin{cases}
1 & \text{if } v_i \neq \hat{v}_i \\
0 & \text{if } v_i = \hat{v}_i
\end{cases}
\label{eq:penalty-iv}
\end{align}
\vspace{+2mm}
The average penalties  $P_S$ and $P_V$ for SFC and VNF identifier, respectively, are calculated as follows: 
\vspace{+5mm}
\begin{equation}
P_S = \frac{1}{N}\sum_{i=1}^{N}\ {P}_{s}^{(i)}
\label{eq:avg-penalty-s}
\end{equation}
\begin{equation}
P_{V} = \frac{1}{N}\sum_{i=1}^{N}\ {P}_{v}^{(i)}
\label{eq:avg-penalty-iv}
\end{equation}
\vspace{+2mm}
  \(\lambda_{ce}\), \(\lambda_s\), and \(\lambda_{v}\) weights are utilized to emphasize/de-emphasize the contribution of cross-entropy loss, SFC and VNF identifier penality components, respectively, thereby controlling their relative impact subject to:
  \vspace{+5mm}
\begin{equation}
\lambda_{ce} + \lambda_s + \lambda_{v} = 1
\label{eq:lambda-sum}
\end{equation}

The total loss \(L\) is then defined as follows:
\vspace{+5mm}
\begin{equation}
L = \lambda_{ce}\, L_{ce} + \lambda_s\, P_S + \lambda_{v}\, P_{v}
\label{eq:overall-loss}
\end{equation}
Where \(L_{ce}\) denotes the standard cross-entropy loss. 

During the training phase, the combined loss function directs the model to minimize the cross-entropy error and accurately predict the SFC and Idle VNF IDs. The model's ability to capture these crucial elements is reinforced by this additional supervision. Future extensions can include incorporation of different reward functions.  All of the parameters in the formulas are summarized in \tablename\hspace{0.1pt}~\ref{tab:notations}.

\section{Performance Analysis}
The experiments were carried out on a system equipped with NVIDIA A100-PCIE-40GB GPUs, each of which features 40GB of memory. The dataset includes schema, 16568 sets of natural-language queries, equivalent SQL queries and ground-truth answers of those queries, which are all manually crafted. It is divided into three parts: 75\% for training, 12.5\% for validation, and 12.5\% for testing.

\begin{table}[!t]
    \centering
    \caption{Configuration Parameters for LiLMs}
    \begin{tabular}{|l|c|}
        \hline
        \textbf{Parameter} & \textbf{Value} \\
        \hline
        Learning Rate & 4e-5 \\\hline
        Batch Size & 2 \\\hline
        Max Length of Tokens & 512 \\\hline
        Epochs & 10 \\\hline
        \(\lambda_{ce}\) & 0.1 \\\hline
        \(\lambda_{v}\) & 0.3 \\\hline
        \(\lambda_{s}\) & 0.6 \\
        \hline
    \end{tabular}
    
    \label{tab:training_params}
    
\end{table}

\begin{table}[!t]
    \centering
    \caption{ \noteblue{Configuration Parameters for SQLCoder}}
    \begin{tabular}{|l|c|}
        \hline
        \textbf{Parameter} & \textbf{Value} \\
        \hline
        Learning Rate & 1e-5 \\\hline
        Batch Size & 2 \\\hline
        Max Length of Tokens & 1024 \\\hline
        r & 16\\\hline
        $\alpha$ & 32 \\\hline
        Epochs & 10 \\
        \hline
    \end{tabular}
    
    \label{tab:training_params2}
    
\end{table}

\begin{figure}[!t]
    \centering
    \begin{subfigure}[b]{0.35\textwidth}
        \centering
        \includegraphics[width=\textwidth, trim=0cm 0cm 0cm 0cm, clip]{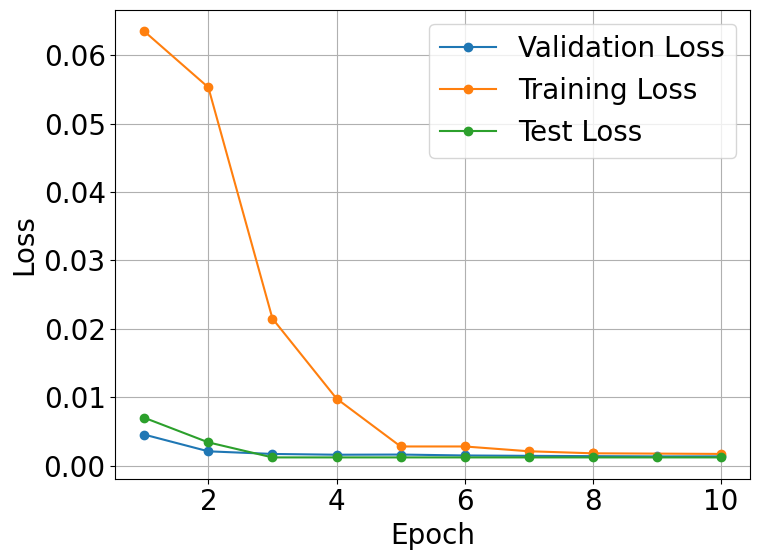}
        \caption{Loss for FLAN-T5}
        \label{fig:losst5}
    \end{subfigure}
    \begin{subfigure}[b]{0.35\textwidth}
        \centering
        \includegraphics[width=\textwidth, trim=0cm 0cm 0cm 0cm, clip]{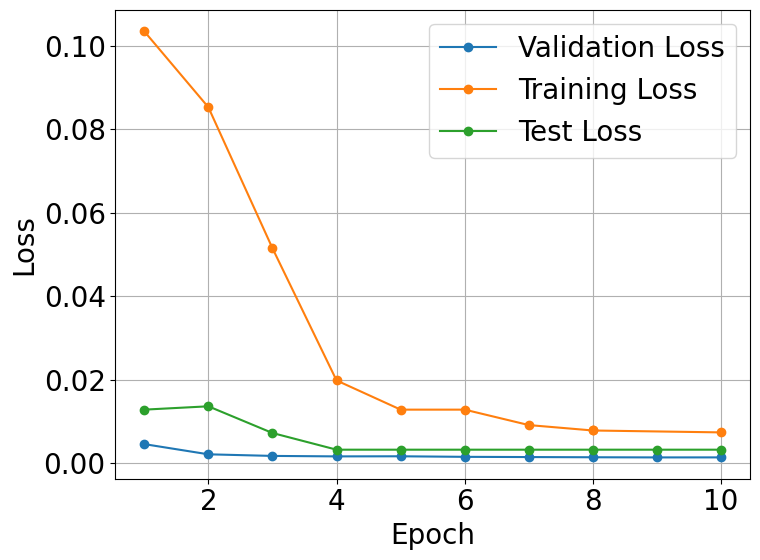}
        \caption{Loss for BART}
        \label{fig:lossbart}
    \end{subfigure}
    \begin{subfigure}[b]{0.35\textwidth}
        \centering
        \includegraphics[width=\textwidth, trim=0cm 0cm 0cm 0cm, clip]{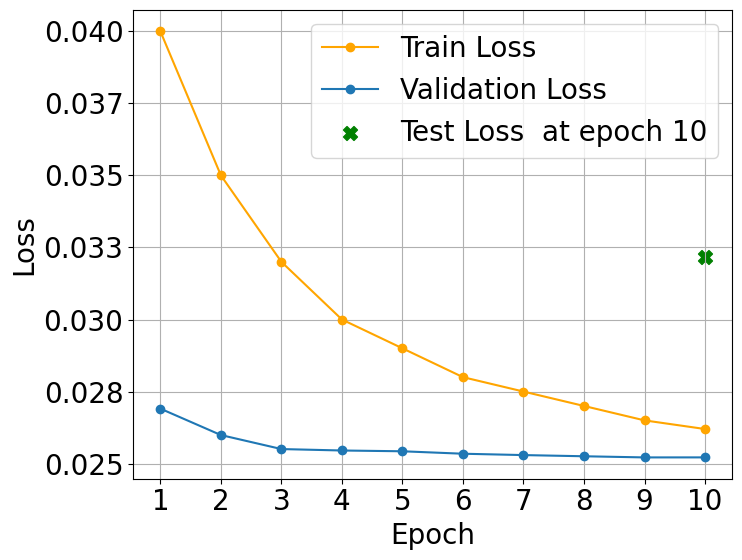}
        \caption{Loss for SQLCoder}
        \label{fig:losssqlcoder}
    \end{subfigure}
    \caption{Training, testing, and validation loss for FLAN-T5, BART, and SQLCoder}
    \label{fig:loss}
\end{figure}

\begin{table*}[!t]
\centering
\caption{Comparison of Metrics for BART, FLAN-T5, and SQLCoder}
\label{tab:comparison_metrics}
\begin{tabular}{|c|c|c|c|}
\hline
\textbf{Metric}         & \textbf{BART}   & \textbf{FLAN-T5} & \textbf{SQLCoder} \\ \hline
Accuracy (\%)           & 80.2\%          & 94.79\%          & 94.54\%           \\ \hline
Correct / Total         & 1661 / 2071     & 1963 / 2071      & 1958 / 2071       \\ \hline
Processing Time         & 2 h 38 min      & 2 h 2 min        & 54 h 43 min       \\ \hline
Perplexity              & 1.0073          & 1.0016           & 1.03              \\ \hline
\end{tabular}
\end{table*}

\begin{figure}[]
    \centering
    \includegraphics[width=0.48\textwidth, trim=0cm 0cm 0cm 0cm, clip]{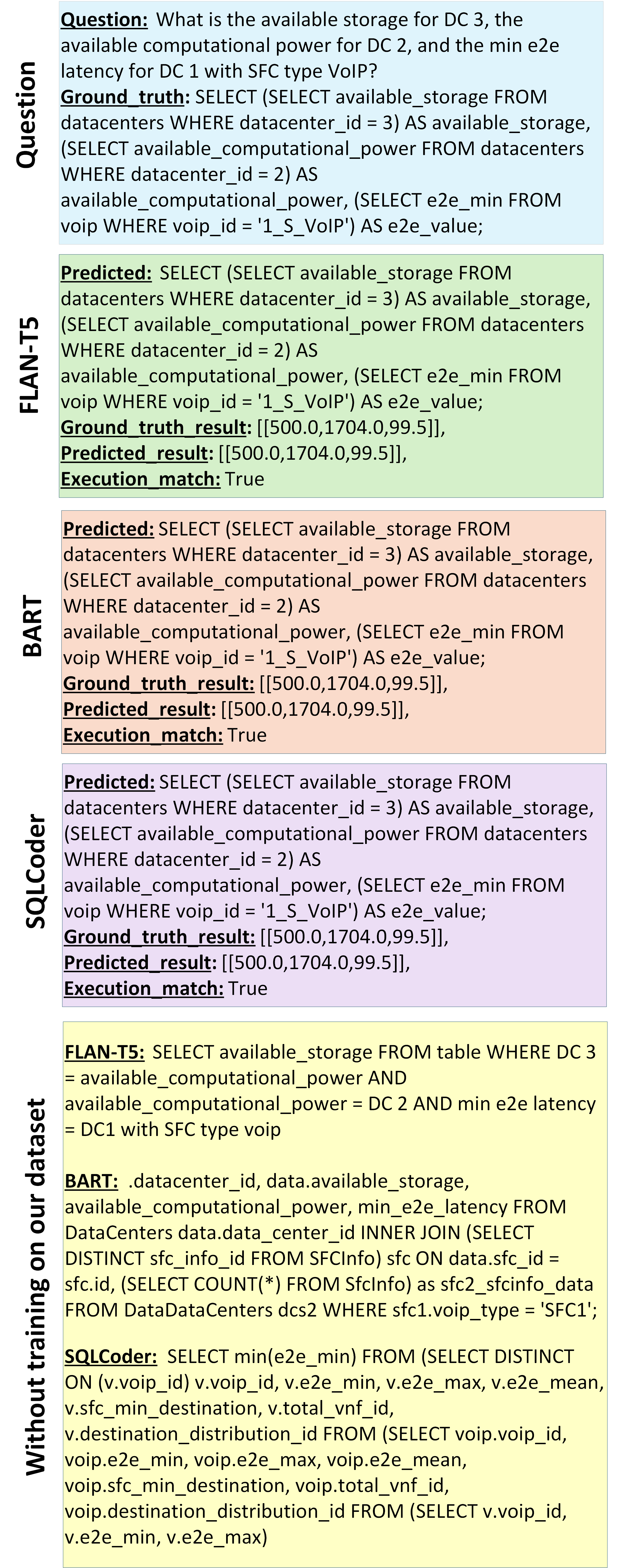}
    \caption{Ground truth and correct predictions for LMs}
    \label{fig:correct}
\end{figure}

 Other configuration parameters are summarized in \tablename\hspace{0.1pt}~\ref{tab:training_params}, which are same for both LiLMs, thereby ensuring that any observed performance differences stem exclusively from their intrinsic architectural properties. In our experiments, SQLCoder is deployed with 8-bit quantization for efficient memory usage and inference, and fine-tuned using LoRA. The LoRA configuration utilizes a rank parameter of r=16, a scaling factor $\alpha =32$, and a dropout rate of 0.05. All of the parameters are summarized in \tablename\hspace{0.1pt}~\ref{tab:training_params2}
All of these values in  \tablename\hspace{0.1pt}~\ref{tab:training_params} and \tablename\hspace{0.1pt}~\ref{tab:training_params2} were carefully chosen after multiple trial runs.

We employ the BART-based model, which has a maximum sequence length of 512 tokens. This model consists of 6 encoder and 6 decoder transformer blocks, each containing 768 hidden dimensions, for a total of around 139 million parameters \cite{bartpaper}. Additionally, we utilize the FLAN-T5-base model, which has a maximum sequence length of 512 tokens. FLAN-T5-base consists of 12 encoder and 12 decoder transformer blocks, each with 768 hidden dimensions and  248 million parameters \cite{chung2022}.
We employ the SQLCoder model from Defog, which is a 15 billion-parameter decoder-only transformer \cite{defog2023sqlcoder}, which is finetuned for our dataset by LoRA.  To further reduce its memory footprint and speed up inference, we quantize the model to 8-bit precision using the bitsandbytes library, enabling deployment with minimal impact on accuracy \cite{bitsandbytes2023}.

Given the maximum input sequence limitation of 512 tokens for both FLAN-T5 and BART, a pre-processing step is utilized to manage input length efficiently. In this step, based on specific keywords identified in the user query, only the relevant portions of the relational database schema are provided to the LiLM. For instance, if the question includes keywords such as "idle VNFs", the schema related to the idle  VNFs table is included. Similarly, if the question pertains to a specific type of SFC, only the schema segments relevant to that SFC type are selected. For combination queries involving multiple aspects, the corresponding schema parts for all mentioned components are included. In cases where no specific keywords are detected, the full schema is presented to the LiLM. This targeted schema selection strategy helps conserve input space, ensuring critical information remains within the model’s token limit.
\begin{figure}[]
    \centering
    \includegraphics[width=0.43\textwidth, trim=0.2cm 0.2cm 0cm 0cm, clip]{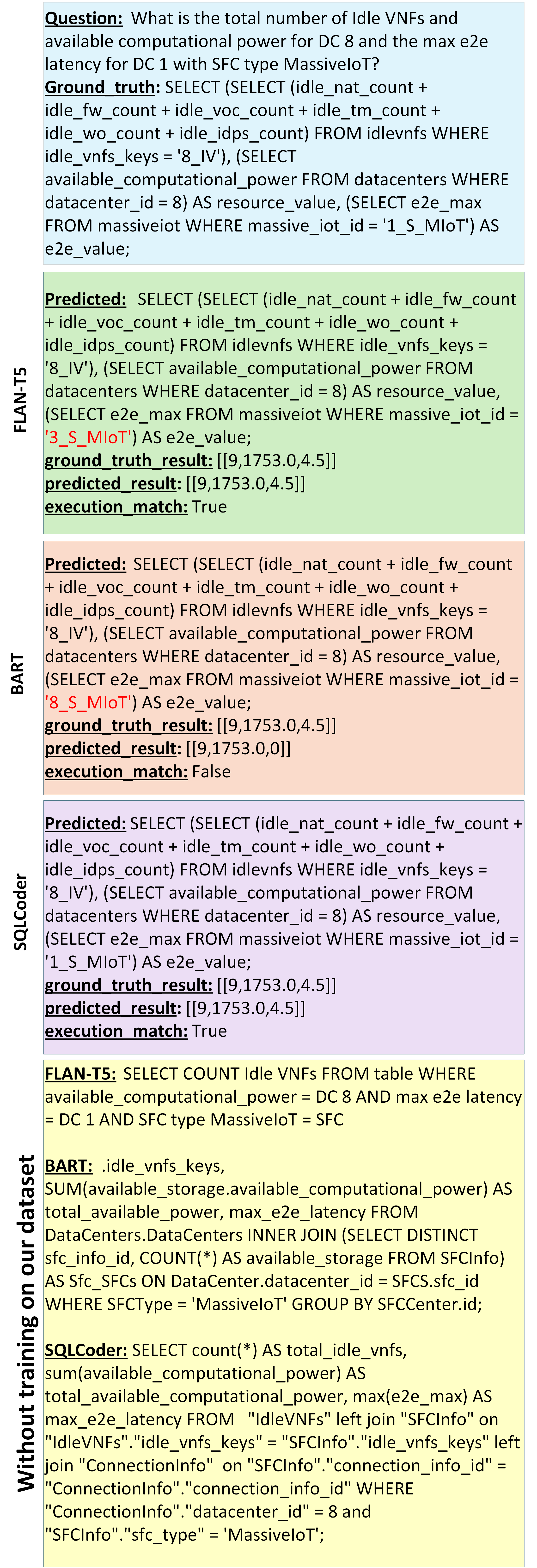}
    \caption{Ground truth and incorrect predictions for LMs}
    \label{fig:incorrect}
\end{figure}

\begin{figure}[]
        \centering
        \includegraphics[width = 0.45\textwidth, trim=0cm 0cm 0cm 0cm,clip]{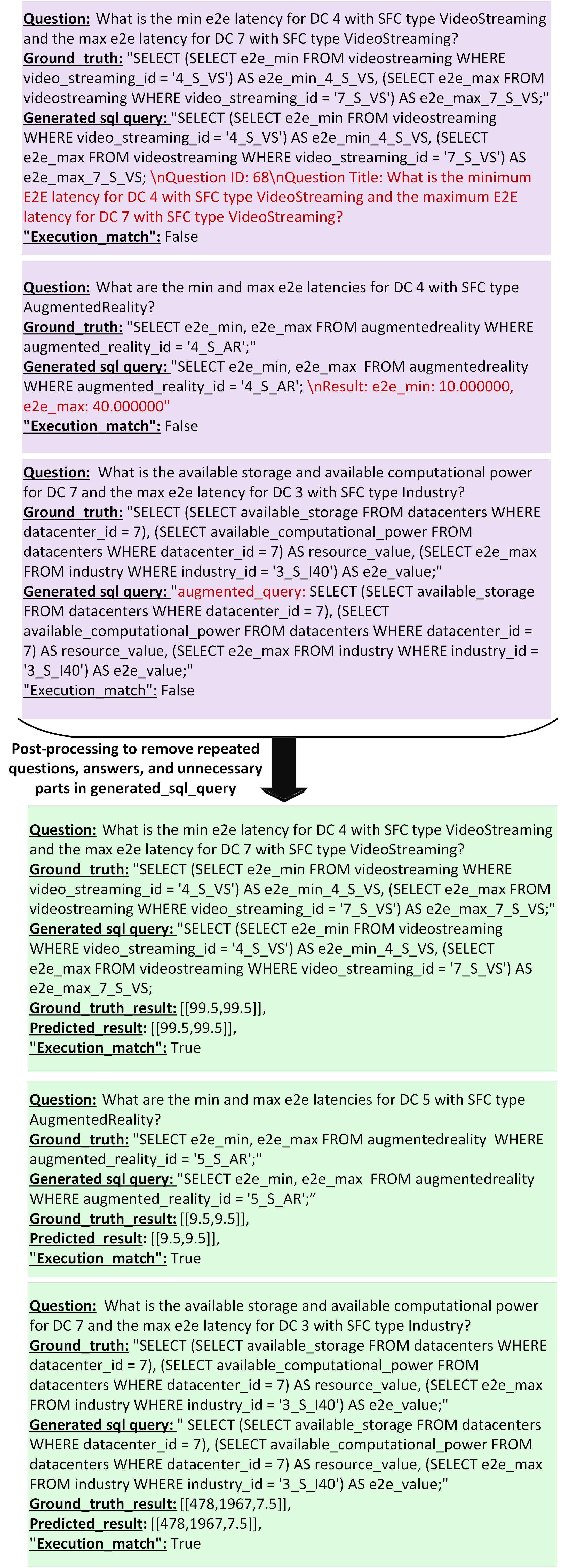}
        \caption{Questions and generated SQL queries for SQLCoder}
        \label{fig:sqlcoder} 
\end{figure}

\figurename \ref{fig:losst5} presents the training, validation, and test loss curves for the FLAN-T5 model over ten epochs. Initially, all losses start at relatively high values and quickly decreasing during the initial epochs. Validation and test losses closely mirror the training loss trend, stabilizing swiftly and indicating the model's strong generalization capabilities. Comparatively, 
\figurename \ref{fig:lossbart} depicts the loss curves for the BART model. The initial losses for BART are higher, but similarly experience rapid decreases within the first few epochs. By the fourth epoch, losses stabilize, with training loss showing the lowest values followed by evaluation and test losses. This rapid and stable convergence suggests effective training and generalization.

\figurename \ref{fig:losssqlcoder}  depicts the training and validation loss curves for the SQLCoder model over ten epochs. The test loss is reported only at epoch 10 due to GPU memory constraints. Similar to LiLMs, training and validation loss of SQLCoder starts at high values and drop during few first epochs  before leveling off around epoch 6. When comparing all three models, all exhibit efficient convergence and robust generalization. However, FLAN-T5 achieves lower overall loss values.

\tablename\hspace{0.1pt}~\ref{tab:comparison_metrics} compares the performance of LiLMs and the LLMs in terms of accuracy, correct predictions, and processing time. The accuracy indicates the proportion of queries generated by the models that match exactly with the correct query word-for-word, thereby retrieving the correct answer from the relational database. FLAN-T5 achieved an accuracy of 94.79\%, resulting in 1963 correct predictions, and completed its computations significantly faster. Conversely, BART reached an accuracy of 80.2\% with 1661 correct predictions but required a considerably longer processing time. Moreover, SQLCoder attains the accuracy of 94.54 \% ( 1958 correct answers), close to FLAN-T5, but with substantial computational cost of 54 hours and 43 minutes of processing.
These results demonstrate FLAN-T5's superior performance in both accuracy and efficiency for our dataset, specifically in generating precise queries necessary for accurate information retrieval from the relational database. Additionally, we calculated perplexity, which measures the average number of choices the model is confused between when predicting the next word/token, that is, 1.0073 for BART, and 1.03 for SQLCoder,  compared to 1.0016 for FLAN-T5, meaning that FLAN-T5 is  more confident in its predictions.

Two illustrative examples of correct and incorrect predictions for the LiLMs and the LLM are presented in \figurename\hspace{0.1pt} \ref{fig:correct} and \figurename\hspace{0.1pt} \ref{fig:incorrect}. In the case of correct predictions, \figurename\hspace{0.1pt} \ref{fig:correct}, all models successfully generate accurate SQL queries, yielding correct execution results. Conversely, in scenarios with incorrect predictions, \figurename\hspace{0.1pt} \ref{fig:incorrect}, FLAN-T5 generates a SQL query with correct  results, albeit identifying an incorrect DC ID. For the same question, BART not only predicts an incorrect DC ID but also generates a SQL query that fails to generate correct answer. In contrast, SQLCoder successfully generates the correct SQL query. Moreover, as illustrated in \figurename\hspace{0.1pt} \ref{fig:correct} and \figurename\hspace{0.1pt} \ref{fig:incorrect}, without additional training, these models consistently produce SQL queries that are syntactically incorrect and cannot be executed on the database. This underscores the limitations of their out-of-the-box capabilities and the necessity of fine-tuning them on domain-specific data.
The generated outputs by SQLCoder are depicted in \figurename\hspace{0.1pt} \ref{fig:sqlcoder}. As shown in \figurename\hspace{0.1pt} \ref{fig:sqlcoder}, SQLCoder sometimes outputs queries containing extra content, such as repeated natural-language questions, answers, or other text fragments. While training SQLCoder is computationally intensive, it nevertheless produces semantically correct SQL statements. By applying a lightweight post-processing step to remove those extra parts, we can recover the correct SQL query itself, resulting in an improvement in the SQLCoder accuracy from 94.54\% to 100\%.


\section{Conclusion}

In this paper, we have proposed LiLM-RDB-SFC, a novel framework integrating a Light Language Model with Relational Database-guided DRL for optimized SFC provisioning. Specifically, the network state information, including final VNF allocations determined by the DRL model, SFC configurations, and DC information, has been utilized by LMs to generate precise SQL queries corresponding to natural-language queries. Querying the current state of SFCs, DCs, and VNFs provides critical insights into real-time resource utilization and potential bottlenecks, significantly enhancing future resource allocation strategies and responsiveness to dynamic network demands.
Our evaluations have shown that the FLAN-T5 model significantly outperforms BART by achieving lower loss values (0.00161 vs 0.00734), higher accuracy (94.79\% vs 80.2\%), shorter processing time (2 h 2 min vs 2 h 38 min) with a higher number of correct query predictions (1963 vs 1661). Compared to SQLCoder, FLAN-T5 has similar accuracy of 94.79\% compared to 94.54\% for SQLCoder, but with 96\% lower processing time.
In the future, different LMs will be utilized with different demand profiles.


\section*{Acknowledgment}
This work is supported by the Natural Sciences and Engineering
Research Council of Canada (NSERC) Alliance Program, MITACS
Accelerate Program, and NSERC CREATE TRAVERSAL program.
\bibliographystyle{IEEEtran}

\end{document}